\documentclass[pra,amsfonts,twocolumn,showpacs]{revtex4}

\usepackage{amsmath}
\usepackage{graphicx}
\usepackage{bm}

\begin{document}

\def\be{\begin{equation}}
\def\ee{\end{equation}}
\def\Tr{\mathop{\rm Tr}}
\def\tr{\mathop{\rm tr}\nolimits}
\def\br{{\bf r}}
\def\bq{{\bf q}}
\def\bk{{\bf k}}
\def\vp{\varphi}
\newcommand{\corr}[1]{\langle #1\rangle}
\newcommand{\ccorr}[1]{\langle\langle #1\rangle\rangle}
\newcommand{\Corr}[1]{\left\langle #1\right\rangle}
\newcommand{\sign}{\mathop{\rm sign}}
\newcommand{\diag}{\mathop{\rm diag}}
\newcommand{\eps}{\varepsilon}
\renewcommand{\Re}{\mathop{\rm Re}}
\renewcommand{\Im}{\mathop{\rm Im}}
\newcommand{\const}{\text{const}}

\def\ff{f} 
\def\ETh{E_{\text{Th}}}

\def\alphaLO{\alpha}
\def\alphaMS{\beta}

\title{Subgap states in disordered superconductors}

\author{M. A. Skvortsov}
\affiliation{L. D. Landau Institute for Theoretical Physics,
142432 Chernogolovka, Russia}
\affiliation{Moscow Institute of Physics and Technology, 141700 Moscow, Russia}

\author{M. V. Feigel'man}
\affiliation{L. D. Landau Institute for Theoretical Physics,
142432 Chernogolovka, Russia}
\affiliation{Moscow Institute of Physics and Technology, 141700 Moscow, Russia}

\date{October 1, 2013}

\begin{abstract}
We revise the problem of the density of states in disordered superconductors.
Randomness of local sample characteristics translates to the quenched spatial
inhomogeneity of the spectral gap, smearing the BCS coherence peak.
We show that various microscopic models of potential and magnetic disorder
can be reduced to a universal phenomenological random order parameter model,
whereas the details of the microscopic description are encoded
in the correlation function of the order parameter fluctuations.
The resulting form of the density of states is generally described by two
parameters: the width $\Gamma$ measuring the broadening of the BCS peak,
and the energy scale $\Gamma_\text{tail}$ which controls the exponential
decay of the density of the subgap states.
We refine the existing instanton approaches for determination
of $\Gamma_\text{tail}$ and show that they appear as the limiting cases
of a unified theory of optimal fluctuations in a nonlinear system.
Application to various types of disorder is discussed.
\end{abstract}



\maketitle

\section{Introduction}

Formation of the superconductive state is intimately related
to the suppression of the quasiparticle density of states (DOS)
in the vicinity of the Fermi energy.
This effect is most pronounced for $s$-wave paring
leading to a hard gap in the quasiparticle spectrum.
If the time-reversal invariance is not broken, the DOS
follows the standard BCS expression,
\be
\label{BCS}
  \rho_\text{BCS}(E) = \rho_0 \Re \frac{E}{\sqrt{E^2-\Delta^2}} ,
\ee
where $\rho_0$ is the normal-metal DOS.
Equation (\ref{BCS}) applies both to clean and disordered
systems~\cite{AG1958,Anderson1959}, indicating that thermodynamics
of superconductors is insensitive to single-particle
dynamics provided that a trajectory has its time-reversed counterpart
needed to form a Cooper pair (Anderson theorem).

\begin{figure}[t]
\includegraphics[width=0.9\columnwidth]{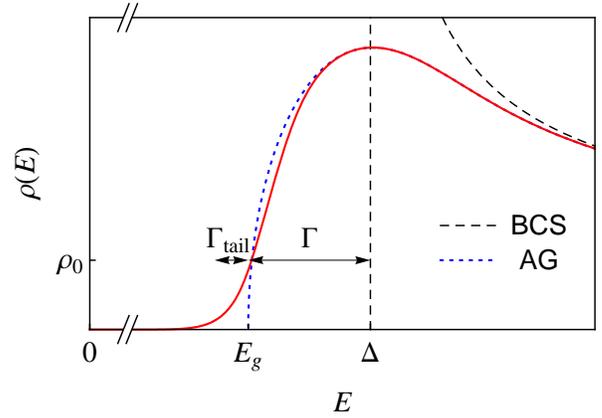}
\caption{Schematic view of the average DOS
in a dirty superconductor (solid line).
Broadening of the BCS peak (dashed line) is mainly described
by the semiclassical approximation (dotted line), with
the full DOS containing a significant tail of the subgap states.}
\label{F:DOS}
\end{figure}

Breaking the time-reversal symmetry
(e.g., by magnetic impurities \cite{AG},
a supercurrent \cite{Anthore-2003},
a magnetic field in small superconducting grains/films \cite{Maki-H})
lowers the critical temperature of the transition and
smears the coherence peak (\ref{BCS}).
Various depairing scenarios are to a large extent equivalent
\cite{Maki-Superconductivity} and can be described by a single
dimensionless parameter
\be
\label{eta-gen}
  \eta = \frac{1}{\tau_\text{dep}\Delta_0} ,
\ee
where $\tau_\text{dep}^{-1}$ is the depairing rate associated
with a particular mechanism of time-reversal symmetry breaking,
and $\Delta_0$ refers to the average value of the order parameter.
According to the general analysis of Abrikosov and Gor'kov (AG) \cite{AG},
the quasiparticle spectrum remains gapful for sufficiently weak
pair breaking, $\eta<1$ (otherwise gapless superconductivity is expected).
A new renormalized gap edge is located at
\be
\label{Eg}
  E_g(\eta) = (1-\eta^{2/3})^{3/2} \Delta_0 ,
\ee
with the DOS vanishing as $\rho(E)\propto(E-E_g)^{1/2}$,
see dotted line in Fig.~\ref{F:DOS}.

In the seminal paper back in 1971, Larkin and Ovchinnikov
have recognized that the BCS-like form of the DOS may be smeared
even if the time-reversal invariance is not broken \cite{LO71}.
They have considered a phenomenological model with a spatially
varying Cooper-channel interaction constant,
$\lambda(\br)=\lambda_0+\delta\lambda(\br)$,
and have shown that short-scale disorder in $\lambda(\br)$
has two effects on the DOS profile:
First, at the mean-field level, it is equivalent to the AG model \cite{AG}
with some effective deparing parameter $\eta$,
therefore leading to the coherence peak smearing,
but still with the hard gap at $E=E_g$.
Second, this hard gap gets also smeared due to optimal fluctuations
of the field $\lambda(\br)$, leading to the Lifshitz-type \cite{ZL,Lif}
tail of $\rho(E)$ in the subgap region, $E<E_g$.

The resulting form of the average DOS in a disordered superconductor
is shown schematically in Fig.~\ref{F:DOS}.
Its main part is given by the AG theory (dotted line), with the
coherent peak smearing controlled by an energy scale $\Gamma=\Delta_0-E_g$.
This region corresponds to uniform configurations of the superconducting
order parameter. On the contrary, the DOS tail at $E<E_g$
originates from the states localized in traps where the order
parameter is lower than its average value $\Delta_0$.
In this regime $\rho(E)$ strongly fluctuates in space,
with the average DOS decaying typically as a stretched exponent at an energy
scale $\Gamma_\text{tail} < \Gamma$:
\be
\label{rho-Gtail}
  \corr{\rho(E)}
  \propto
  \exp\left[-\left(\frac{E_g-E}{\Gamma_\text{tail}}\right)^\zeta\right] .
\ee

Appearance of sufficiently deep traps is a rear event
which is naturally identified with an instanton
in the quasiclassical equations of superconductivity.
Working in the dirty limit and studying optimal fluctuations
in the Usadel equation \cite{Usadel},
Larkin and Ovchinnikov \cite{LO71} have calculated the average subgap DOS,
\be
\label{subgap-LO}
  \corr{\rho(E)}_\text{LO}
  \propto
  \exp\left( - \alphaLO_d(\eta) \,
  \frac{\Delta_0^2\xi^{d}}{\ff(0)} \, \eps^{(8-d)/4} \right) ,
\ee
which behaves as a stretched exponent of the dimensionless distance $\eps$
from the gap edge,
\be
\label{eps-def}
  \eps = \frac{E_g-E}{E_g} ,
\ee
with the power $\zeta_\text{LO}=(8-d)/4$ dependent on the space
dimensionality $d$ \cite{com-d}. In Eq.~(\ref{subgap-LO}),
$\Delta_0$ is the average value of the order parameter,
$\xi=\sqrt{D/2\Delta_0}$ is the superconducting coherence length
($D$ is the diffusion coefficient), $f(0)$ is the zero Fourier
harmonics of the correlation function of the order parameter
fluctuations induced by quenched disorder in $\lambda(\br)$
[see Eq.~(\ref{f(q)}) below], and $\alphaLO_d(\eta)$ is a function
of the dimensionless depairing parameter $\eta$
[see Eqs.~(\ref{eta}) and (\ref{alpha-d}) below].

The power of $\eps$ in Eq.~(\ref{subgap-LO}) can be easily understood
within the optimal fluctuation approach. Near the AG threshold, at $E\to E_g$,
the system is characterized by a diverging length scale
$L_E \sim \xi \eps^{-1/4}$ \cite{LO71}. To have a quasiparticle state
with an energy $E=(1-\eps)E_g$ below the mean-field gap
one has to locally reduce the order
parameter by an amount of $\delta\Delta\sim\eps\Delta_0$ in a volume
specified by the length $L_E$. The price one has to pay for such
an optimal fluctuation scales as $(\delta\Delta)^2 L_E^d \sim \eps^{2-d/4}$,
in accordance with the result (\ref{subgap-LO}).

Precisely the same model of a fluctuating Cooper constant \cite{LO71}
in a dirty superconductor has been reanalyzed thirty years
later by Meyer and Simons \cite{MeyerSimons2001}
in the framework of the nonlinear $\sigma$ model approach.
Using the instanton analysis
of the $\sigma$ model, they have obtained a somewhat different
optimal fluctuation leading to a different result for the tail
of the subgap states:
\be
\label{subgap-MS}
  \corr{\rho(E)}_\text{MS}
  \propto
  \exp\left( -\alphaMS_d(\eta) \, g_\xi \, \eps^{(6-d)/4} \right) ,
\ee
which is also a stretched exponent but with a different power $\zeta_\text{MS}=(6-d)/4$.
Besides that, the instanton action of Meyer and Simons does not depend
on the order-parameter correlation function $f(\br)$. Instead,
it contains some function $\alphaMS_d(\eta)$ of the depairing
parameter $\eta$ [see Eq.~(\ref{beta-d}) below] and the dimensionless
(in units of $e^2/h$) conductance $g_\xi$ of the region of size $\xi$:
\be
\label{gxi}
  g_\xi
  = 4\pi \nu D \xi^{d-2}
  = 8\pi \nu \Delta_0 \xi^{d} .
\ee
Appearance of the conductance $g_\xi$ in the exponent of Eq.~(\ref{subgap-MS})
indicates that this expression cannot be obtained at the level
of the saddle-point (Usadel) equation but requires
the usage of the full nonlinear field theory.

Expression (\ref{subgap-MS}) for the density of subgap states has been
obtained for a variety of disordered superconducting systems
\cite{MeyerSimons2001,LamacraftSimons,Vavilov2001,OSF01,MarchettiSimons},
where the semiclassical approximation predicts a square-root vanishing
of the DOS, $\rho(E)\propto\sqrt{E-E_g}$.
In particular, it was observed in hybrid normal-metal -- superconductor (NS)
systems \cite{Vavilov2001,OSF01}, and in bulk superconductors with
magnetic impurities \cite{LamacraftSimons,MarchettiSimons}.
Mathematically it bears a close analogy with the Tracy-Widom
distribution for the DOS tail in the Random matrix theory (RMT) \cite{TracyWidom},
generalizing it from $d=0$ to an arbitrary dimensionality $d$.
Based on these findings it is widely believed that Eq.~(\ref{subgap-MS})
provides a universal description of the subgap DOS tail in disordered
superconductors. However the discrepancy with the analysis of Larkin
and Ovchinnikov existing at least for the model of the random Cooper
channel constant still remains unresolved.

The purpose of this paper is to fill this gap by clarifying the origin
of the two types of instantons discussed in Refs.~\onlinecite{LO71} and \onlinecite{MeyerSimons2001}
[leading to Eqs.~(\ref{subgap-LO}) and (\ref{subgap-MS})].
We will show that they correspond to different limits of \emph{a unique instanton
solution} realized for small and large $\eps$, respectively.
Hence, the Larkin-Ovchinnikov instanton can be continuously deformed
into the Meyer-Simons instanton by changing the distance to the gap, $\eps$.
Such an unusual situation is a consequence of the nonlinearity of the
Usadel equation. Therefore averaging over the random order parameter field $\Delta(\br)$
produces a nonlinear term \cite{ZL} which will compete with the intrinsic
nonlinearity of the problem.
This should be contrasted with the problem of fluctuation bound states in the
Schr\"odinger equation with random potential \cite{ZL,Lif}, where the only
source of nonlinearity is due to averaging over disorder.

The paper is organized as follows.
In Sec.~\ref{S:ROP} we introduce the random order parameter (ROP) model
and derive its effective action in the large-scale limit.
In Sec.~\ref{S:instantons} we analyze the instanton solutions
with the broken replica symmetry and recover the results
(\ref{subgap-LO}) and (\ref{subgap-MS}) in different limits.
The summary and applications of the ROP model are discussed
in Sec.~\ref{S:ROP-summary}.
Gap smearing in superconductors with magnetic impurities is reconsidered
in Sec.~\ref{S:magnetic}. We conclude with discussion of the results
obtained in Sec.~\ref{S:discussion}.
Technical details are relegated to Appendix.

\section{Random order parameter model}
\label{S:ROP}

\subsection{Model}

We start with the simplest example when the gap smearing results from
quenched inhomogeneity in the pairing potential,
\be
\label{d=d+d}
  \Delta(\br)=\Delta_0+\Delta_1(\br),
\ee
which is assumed to be a real
Gaussian random field specified by the correlation function
\be
\label{f-def}
  \corr{\Delta_1(\br) \Delta_1(\br')} = f(\br-\br') .
\ee
The function $f(r)$ is supposed to be short-ranged,
with the correlation length, $r_c$, being smaller than
the superconducting coherence length \cite{com-rc}:
\be
\label{rc<xi}
  r_c<\xi.
\ee
The superconductor is assumed to be in the dirty limit,
$T_c\tau\ll1$, where $\tau$ is the elastic scattering time.

The main simplification of this model, that will be referred to as
the random order parameter (ROP) model, is that $\Delta(\br)$ is considered
as a given external field which should not be determined self-consistently.

The phenomenological ROP model universally emerges as an intermediate step
in studying various types of disorder in the \emph{singlet} case,
when spin effects can be neglected \cite{LO71,MeyerSimons2001,FS2012}
(a more general situation will be considered in Sec.~\ref{S:magnetic}).
The function $f(\br)$ in Eq.~(\ref{f-def}) then bears information
on the original inhomogeneity in a particular microscopic model,
see Sec.~\ref{SS:ROP-apps}.

\subsection{Sigma model}
\label{SS:RSM}

The ROP model was treated by Larkin and Ovchinnikov \cite{LO71}
in terms of equations of motion (Usadel equation),
and by Simons and co-authors \cite{MeyerSimons2001,LamacraftSimons,MarchettiSimons}
within the nonlinear $\sigma$-model formulation.
Aiming to compare the two approaches, we choose to work in the functional
language of the diffusive $\sigma$ model.
To study the DOS in a field of a given $\Delta(\br)$
at a particular energy $E$ one can use either
its supersymmetric or replica version. We prefer to deal with
the real-energy replica $\sigma$ model formulated in terms of the field
$Q(\br)$ acting in the direct product of the replica, Nambu and spin spaces
(the latter is redundant in the singlet case considered but will be
employed for the study of magnetic impurities in Sec.~\ref{S:magnetic})
\cite{LamacraftSimons,Efetov,Finkelstein-review,AST,we-spinflip}.

Choosing the order parameter to be real, we write the $\sigma$-model
action as
\be
  S
  =
  \frac{\pi\nu}{4}
  \int d\br \,
  \tr \left[
    D (\nabla Q)^2 + 4 (iE\tau_3 - \Delta(\br)\tau_1) Q
  \right]
  ,
\ee
where $\tau_i$ are Pauli matrices in the Nambu space.

Averaging over quenched disorder in $\Delta(\br)$ with the help
of Eqs.~(\ref{d=d+d}) and (\ref{f-def}), we arrive at the action
for the field $Q$:
\be
\label{S01}
  S = S_0 + S_\text{dis} ,
\ee
where
\begin{gather}
\label{S0}
  S_0
  =
  \frac{\pi\nu}{4}
  \int d\br \,
  \tr \left[
    D (\nabla Q)^2 + 4 (iE\tau_3 - \Delta_0\tau_1) Q
  \right] ,
\\
  S_\text{dis}
  =
  -
  \frac{(\pi\nu)^2}{2}
  \int d\br \, d\br' \,
  f(\br-\br') \,
  \tr[\tau_1 Q(\br)]
  \tr[\tau_1 Q(\br')] .
\label{Sdis}
\end{gather}

\subsection{Effective long-wavelength action}
\label{SS:LRA}

The term $S_\text{dis}$ [Eq.~(\ref{Sdis})] contains an additional trace
in the replica space and therefore does not contribute to the replica-diagonal
saddle-point equation of motion. According to
Larkin and Ovchinnikov \cite{LO71}, in order to see effects of disorder
at the saddle-point level in the long-wavelength limit (with momenta $q<q_0$),
one has to average $S_\text{dis}$ over fast
fluctuations of the field $Q$ (cooperons and diffusons).
This procedure generates an effective depairing term
\cite{MeyerSimons2001,we-spinflip}
\be
\label{Seta}
  S_\eta
  =
  - \frac{\pi\nu\Delta_0\eta}{4}
  \int d\br \,
  \tr ( \tau_3 Q )^2
  ,
\ee
where the coefficient $\eta$ is expressed in terms of the Fourier transform
of the order-parameter correlation function,
\be
\label{f(q)}
  f(\bq) = \corr{\Delta_1 \Delta_1}_{\bq} ,
\ee
as
\be
\label{eta}
  \eta
  =
  \frac{2}{\Delta_0}
  \int
  \frac{\ff(\bq)}{Dq^2}
  \frac{d^d\bq}{(2\pi)^d}
  .
\ee
In this derivation it was assumed that the regions of large
momenta ($q>q_0$) contributing to Eq.~(\ref{eta}) and small momenta ($q<q_0$)
for which we derive an effective theory are well separated.
This is true in 3D \cite{LO71}, marginally true in 2D \cite{FS2012}
and wrong in 1D, see Sec.~\ref{SS:dim} for details.

Having eliminated short-range degrees of freedom we end up
with an effective long-range ($r\gg r_c$) action for the field $Q$:
\be
\label{S3}
  S = S_0 + S_\eta + S_\text{dis} ,
\ee
where $S_0$ is given by Eq.~(\ref{S0}), $S_\eta$ is given by Eq.~(\ref{Seta}),
and $S_\text{dis}$ can be written in the local form:
\be
\label{Sdis-local}
  S_\text{dis}
  =
  -
  \frac{(\pi\nu)^2}{2} \ff(0)
  \int d\br \,
  [\tr\tau_1 Q(\br) ]^2
  .
\ee

At this stage we may trace the difference between the approaches
of Refs.~\cite{LO71} and \cite{MeyerSimons2001}.
In order to reproduce the analysis of Larkin and Ovchinnikov \cite{LO71}
one has to decouple the term $S_\text{dis}$ [Eq.~(\ref{Sdis-local})]
with the Gaussian white-noise order parameter field $\Delta_1(\br)$
and treat the resulting problem in the saddle-point approximation
assuming the solution is replica symmetric.
As we will see in Sec.~\ref{SS:LO}, in terms of the $Q$-only action (\ref{S3})
this corresponds to instanton solutions with infinitesimally
small replica symmetry breaking.
On the other hand, Meyer and Simons \cite{MeyerSimons2001} did not use
the saddle-point approximation but completely neglected the term $S_\text{dis}$
which accounts for long-range fluctuations of the order-parameter field.
Their instanton solution originating from the nonlinearity of the underlying
field theory has a nontrivial replica structure discussed in Sec.~\ref{SS:MS}.
Below we will analyze the action (\ref{S3}) and clarify the validity
of approximations employed in Refs.~\cite{LO71} and \cite{MeyerSimons2001}.

\section{Optimal fluctuations in a non-linear system}
\label{S:instantons}

\subsection{Saddle-point equations}

Here we analyze the saddle points of the action (\ref{S3})
which have the replica-diagonal form:
\be
\label{Q0}
  (Q_0)^{ab}
  =
  \delta_{ab}
  \left[
    \tau_3 \cos\theta^a
  + \tau_1 \sin\theta^a
  \right]
  ,
\ee
where Latin indices refer to the replica space,
and the spectral angle $\theta^a(E)$ depends on the energy considered.

The simplest is the \emph{replica-symmetric}\/
saddle point, with $\theta^a=\theta_0$ for all $a=1,\dots,n$.
For a replica-symmetric solution, the actions $S_0$ and $S_\eta$
are proportional to the number of replicas, $n$, whereas the action $S_\text{dis}$
is proportional to $n^2$ and does not contribute to the saddle-point
(Usadel) equation in the replica limit $n\to0$.
Then the saddle-point equation for a uniform $Q_0$ immediately
reproduces the AG equation for the spectral angle in the model
of magnetic impurities \cite{AG}:
\be
\label{Usadel-AG}
  iE \sin\theta_0
+ \Delta_0 \cos\theta_0
  -
  \Delta_0 \eta \cos\theta_0 \sin\theta_0
  =
  0
  .
\ee
The corresponding DOS, $\rho(E) = \rho_0 \Re \cos\theta_0$,
characterized by the hard gap at $E_g$ [Eq.~(\ref{Eg})]
is shown by the dotted line in Fig.~\ref{F:DOS}.

The subgap states are associated with localized saddle-point solutions
\emph{with broken replica symmetry} \cite{MeyerSimons2001}.
To study them it is convenient to write \cite{LamacraftSimons,OSF01}
\be
  \theta^a=\pi/2+i\psi^a .
\ee
The replica-symmetric mean-field solution, $\psi^a=\psi_0$,
is real for $E<E_g$, and appearance of a finite DOS is related
to configurations with complex $\psi$.

The set of the saddle-point equations for the action (\ref{S3})
in terms of the angle $\psi$ takes the form:
\be
\label{SP}
  - \xi^2 \nabla^2\psi^a
  + F(\psi^a)
  +
  \kappa
  \sinh\psi^a
  \sum_{b}
  \cosh\psi^b
  =
  0 ,
\ee
where $\xi$ is the coherence length, $\xi^2=D/2\Delta_0$,
the parameter $\kappa$ is related to the low-momentum correlator
of the order parameter fluctuations:
\be
  \kappa
  =
  \frac{4\pi\nu\ff(0)}{\Delta_0}
  =
  \frac{g_\xi}{2} \frac{\ff(0)}{\Delta_0^2\xi^d},
\ee
and the function $F(\psi)$ is given by
\be
\label{F(psi)-def}
  F(\psi)
  =
  - \frac{E}{\Delta_0} \cosh\psi
  + \sinh\psi
  -
  \eta
  \sinh\psi
  \cosh\psi .
\ee
Equation $F(\psi)=0$ is equivalent to the AG equation (\ref{Usadel-AG}).

Behavior of the function $F(\psi)$ for real arguments depends on the
relation between $E$ and $E_g$.
Below the gap ($E<E_g$), equation $F(\psi)=0$ has two solutions:
$\psi_0$ (AG solution) and $\psi'>\psi_0$. They merge at $E=E_g$,
where an analytic solution is possible, yielding
\be
\label{cosh-g}
  \cosh\psi_g=\eta^{-1/3}.
\ee
Above the gap ($E>E_g$), equation $F(\psi)=0$ has no real solutions.
For small deviation from the gap, $E\to E_g$, the function $F(\psi)$
can be expanded near the AG solution as
\be
\label{Omega-rho}
  F(\psi) \approx \Omega (\psi-\psi_0) - \rho (\psi-\psi_0)^2
  ,
\ee
with the dimensionless parameters
\be
  \Omega = (1-\eta^{2/3}) \sqrt{6\eps}
  ,
\qquad
  \rho = \frac32 \eta^{1/3} \sqrt{1-\eta^{2/3}}
  ,
\ee
where $\eps$ is defined in Eq.~(\ref{eps-def}).

\subsection{Instantons with broken replica symmetry}
\label{SS:ROP-1step}

Now we concentrate on solutions of Eqs.~(\ref{SP})
with the simplest nontrivial structure in the replica space
\cite{MeyerSimons2001}:
\be
  \psi^a(\br)
  =
  \begin{cases}
    \psi_1(\br), & a=1; \\
    \psi_2(\br), & a=2,\dots,n.
  \end{cases}
\ee
Such a solution is characterized by two functions, $\psi_1(\br)$
and $\psi_2(\br)$, which satisfy the system of two coupled nonlinear
equations (\ref{SP}), where
\be
  \sum_b\cosh\psi^b
  =
  \cosh\psi_1-\cosh\psi_2
\ee
in the replica limit ($n\to0$).

The system (\ref{SP}) simplifies in the vicinity of the gap edge,
$E\to E_g$,
where variations of $\psi_1(\br)$ and $\psi_2(\br)$ are small
and the replica-mixing term may be linearized.
To write the resulting equations in a dimensionless form,
we measure distance in units of the divergent length scale \cite{LO71}
\be
\label{LE}
  L_E
  =
  \frac{\xi}{\sqrt{\Omega}}
  \sim
  \xi \left( \frac{E_g}{E_g-E} \right)^{1/4}
\ee
and write
\be
\label{psi-phi}
  \psi_{1,2}(\br)=\psi_0+(\Omega/\rho)\phi_{1,2}(\br).
\ee
As a result, we arrive at the following system:
\begin{subequations}
\label{SP12}
\begin{gather}
\label{SP1}
  - \nabla^2\phi_1
  + \phi_1 - \phi_1^2
  = K(\eps) (\phi_2-\phi_1)
  ,
\\
\label{SP2}
  - \nabla^2\phi_2
  + \phi_2 - \phi_2^2
  = K(\eps) (\phi_2-\phi_1)
  .
\end{gather}
\end{subequations}
The replica mixing is controlled by the the single dimensionless
parameter $K(\eps)$:
\be
  K(\eps)
  = \sqrt{\frac{\eps_*}{\eps}}
  ,
\ee
where the energy scale $\eps_*$ given by
\be
\label{eps*}
  \eps_*
  =
  \frac{\kappa^2}{6\eta^{4/3}}
  =
  \frac{g_\xi^2}{24\eta^{4/3}} \left( \frac{\ff(0)}{\Delta_0^2\xi^d} \right)^2
  .
\ee

Sufficiently close to the gap edge, at $\eps\lesssim\eps_*$,
the parameter $K(\eps)$ is large and equations
for $\phi_1(\br)$ and $\phi_2(\br)$ are strongly coupled.
Small values of $K(\eps)$ can be realized only
for large deviations from the gap, at $\eps\gtrsim\eps_*$.

With the exponential accuracy the subgap DOS is determined
by the instanton action:
\be
\label{rho-S}
  \corr{\rho(E)}
  \propto
  \exp
  \left(
  - \gamma_d(\eta) \, g_\xi \, \eps^{(6-d)/4} S_0[K(\eps)]
  \right)
  ,
\ee
where $g_\xi$ given by Eq.~(\ref{gxi}) is the dimensionless
conductance of the region of size $\xi$,
\be
\label{beta}
  \gamma_d(\eta)
  =
  \frac{4}{3} \, 6^{(2-d)/4} \,
  \frac{(1-\eta^{2/3})^{2-d/2}}{\eta^{2/3}} ,
\ee
and $S_0(K)$ is the dimensionless instanton action:
\be
\label{S12}
  S_0(K)
  =
  \frac16
  \int (\phi_2^3 - \phi_1^3) \, d\br .
\ee

Note that the energy dependence of the average DOS (\ref{rho-S})
comes both from the factor $\eps^{(6-d)/4}$ and the energy dependence
of the parameter $K(\eps)$.
Below we will analyze solutions of Eqs.~(\ref{SP12}) in the limiting
cases of small and large values of $K$ and identify them with the
Meyer-Simons and Larkin-Ovchinnikov instantons, respectively.

\subsection{Instanton in the limit $K\to0$}
\label{SS:MS}

In the limit $K\to0$, Eqs.~(\ref{SP12}) decouple
yielding a single equation
\be
\label{Usadel0}
  - \nabla^2 \phi + \phi - \phi^2 = 0
\ee
both for $\phi_1(\br)$ and $\phi_2(\br)$.
This equation has three solutions: two constant solutions,
$\phi(\br)=0$ (corresponding to the AG solution) and $\phi(\br)=1$,
and a spherically symmetric bounce $\phi_\text{inst}^{(d)}(r)$
vanishing for $r\to\infty$. The bounce solution of Eq.~(\ref{Usadel0})
is known explicitly in the 1D geometry:
\be
\label{phi1D}
  \vp_\text{inst}^{(1)}(r)
  =
  \frac{3}{2\cosh^{2}(r/2)} ,
\ee
and can be obtained numerically for other dimensionalities.
The instanton action is determined by the number
\be
\label{a-d}
  s_d
  \equiv
  \frac16
  \int [\vp_\text{inst}^{(d)}(r)]^3 d\br
  =
  \begin{cases}
    6/5, & d=1 , \\
    7.75, & d=2 , \\
    43.7, & d=3 .
  \end{cases}
\ee

To minimize the action (\ref{S12}) we take the trivial AG
solution $\phi_1(\br)=0$ for the first replica and choose
a bounce solution, $\phi_2(\br) = \vp_\text{inst}^{(d)}(r)$,
for the other replicas.
Hence, $S_0(0)=s_d$ and Eq.~(\ref{rho-S})
reproduces the result (\ref{subgap-MS})
of Refs.~\onlinecite{MeyerSimons2001} and \onlinecite{LamacraftSimons}
with
\be
\label{beta-d}
  \beta_d(\eta) = s_d \gamma_d(\eta) .
\ee

\subsection{Instanton for $K\to\infty$, optimal fluctuation,\\
and dimensional reduction}
\label{SS:LO}

In the limit $K\to\infty$, the last terms in Eqs.~(\ref{SP12})
render $\phi_1(\br)$ and $\phi_2(\br)$ nearly equal.
So we may expand their difference in powers of $K^{-1}$ and write
\be
  \phi_1(\br) = \phi(\br),
\qquad
  \phi_2(\br) = \phi(\br) + K^{-1} \chi(\br) + \dots
\ee
Substituting this expansion into Eqs.~(\ref{SP12})
we get
\begin{subequations}
\label{SP12chi}
\begin{gather}
\label{SP1chi}
  - \nabla^2\phi
  + \phi - \phi^2
  = \chi
  ,
\\
\label{SP2chi}
  - \nabla^2\chi
  + \chi - 2\phi\chi
  = 0
  .
\end{gather}
\end{subequations}
Now excluding $\chi(\br)$, we come to the fourth-order
differential equation for the function $\phi(\br)$:
\be
\label{SP-LO}
  [- \nabla^2 + 1 - 2\phi]
  [- \nabla^2\phi + \phi - \phi^2]
  =
  0
  .
\ee

Equation (\ref{SP-LO}) naturally appears in the study of optimal
fluctuations in a nonlinear equation
\be
  -\nabla^2\phi + F(\phi) = h(\br) ,
\ee
where $F(\phi)=\phi-\phi^2$ and $h(\br)$ is a Gaussian $\delta$-correlated
random field \cite{LO71}.
Optimal fluctuation arguments \cite{ZL,Lif} lead to the minimization
of the functional $\int h^2 d\br = \int [-\nabla^2\phi + F(\phi)]^2 d\br$,
and hence to the saddle-point equation
\be
\label{Usadel2}
  [ -\nabla^2 + F'(\phi) ]
  [ -\nabla^2\phi + F(\phi) ] = 0 ,
\ee
coinciding with Eq.~(\ref{SP-LO}). Thus we see that in the replica
formalism the role of the random field $h(\br)$ is played
by the mismatch of solutions for different replicas:
$\chi(\br)\propto\phi_2(\br)-\phi_1(\br)$.

Note that those $\phi(\br)$ solving Eq.~(\ref{Usadel0})
also solve Eq.~(\ref{SP-LO}). However none of them
correspond to an optimal fluctuation since they
have $h=0$ and do not lead to a finite DOS.
Therefore one has to look for another solution
of Eq.~(\ref{SP-LO}).
Quite surprisingly, for the spherically symmetric solutions $\phi(r)$,
there exists an identity valid for an arbitrary function $F(\phi)$
and arbitrary $d$ \cite{Silva}:
\begin{multline}
\label{identity}
  \bigl[ - \Delta^{(d)}_\text{rad} + F'(\phi) \bigr]
  \bigl[ - \Delta^{(d)}_\text{rad}\phi + F(\phi) \bigr]
\\{}
  =
  \bigl[ - \Delta^{(d+2)}_\text{rad} + F'(\phi) \bigr]
  \bigl[ - \Delta^{(d-2)}_\text{rad}\phi + F(\phi) \bigr]
  ,
\end{multline}
where $\Delta^{(d)}_\text{rad}$ is the radial part of the Laplace operator
in $d$ dimensions:
\be
  \Delta^{(d)}_\text{rad}
  =
  \frac{1}{r^{d-1}}\frac{\partial}{\partial r}r^{d-1}
  \frac{\partial}{\partial r} .
\ee

Thus there is a kind of a dimensional reduction:
a nontrivial optimal fluctuation in $d$ dimensions solving Eq.~(\ref{SP-LO})
is just the bounce solution of Eq.~(\ref{Usadel0}) in $d-2$ dimensions.
Note that a somewhat similar dimensional reduction has been obtained
in Ref.~\onlinecite{Parisi-Sourlas-1979}
for the critical behavior of spin systems in a random magnetic field.

The instanton action (\ref{S12}) is given by
\be
\label{SK-LO}
  S_0(K)
  = \frac{1}{2K} \int \phi^2 \chi \, d\br
  = \frac{1}{2K} \int \chi^2 \, d\br ,
\ee
where the last relation follows from Eqs.~(\ref{SP12chi}).
In the optimal fluctuation language, $\exp[-S(K)]$ is just
the probability density for the Gaussian random field $\chi(\br)$.
Substituting Eq.~(\ref{SK-LO}) into the general expression (\ref{rho-S}),
we arrive at the result (\ref{subgap-LO}), where
\be
\label{alpha-d}
  \alpha_d(\eta)
  =
  16\cdot 6^{-d/4} c_d \,
  (1-\eta^{2/3})^{2-d/2} ,
\ee
and $c_d=\lim_{K\to\infty} KS_0(K)$ is the $d$-dependent constant:
\be
\label{c-d}
  c_d
  = 2 \int
  \biggl(\frac{\partial\vp^{(d-2)}_\text{inst}(r)}{\partial r}\biggr)^2
  \frac{d\br}{r^2}
  =
  \begin{cases}
    0.266, & d=1 , \\
    2.09, & d=2 , \\
    24\pi/5, & d=3 .
  \end{cases}
\ee

In the 3D case and at $\eta\ll1$, Eq.~(\ref{subgap-LO}) coincides with
the result of Ref.~\onlinecite{LO71} (where only this limit was considered).
Thus our instantons in the limit $K\to\infty$ directly correspond
to the optimal fluctuations of Larkin and Ovchinnikov,
and the dimensional reduction (\ref{identity}) explains why
did they manage to find an explicit analytic expression
for the optimal fluctuation in the 3D case:
$\phi^{(3)}(r)=\vp_\text{inst}^{(1)}(r)$,
with the latter given by Eq.~(\ref{phi1D}).

\begin{figure}
\includegraphics[width=0.95\columnwidth]{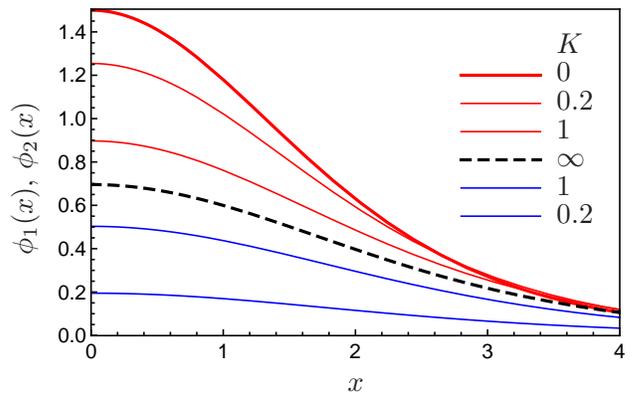}
\caption{Numerical solutions of Eqs.~(\ref{SP12}) in the 1D geometry
($d=1$) for various values of the replica-mixing parameter $K$.
For $K=0$, $\phi_1(x)=0$ and $\phi_2(x)=\vp^{(1)}_\text{inst}(x)$
(upper thick curve).
Solutions for intermediate values ($K=0.2$ and $K=1$)
are shown by the thin solid lines.
In the limit $K\to\infty$, $\phi_1(x)$ and $\phi_2(x)$ approach the asymptote
$\vp^{(-1)}_\text{inst}(x)$ (thick dashed line).}
\label{F:instantons}
\end{figure}

\subsection{Intermediate values of $K$}

Two types of instantons analyzed above
continuously interpolate between each other with variation of $K$.
As an example we show in Fig.~\ref{F:instantons} numerical solutions
of 1D equations (\ref{SP12}) for some intermediate values of $K$.
For small but finite $K$, solutions $\phi_1(r)$ and $\phi_2(r)$
start to deviate from 0 and $\vp^{(d)}_\text{inst}(r)$, respectively,
moving towards each other. Already at $K=1$ they are close,
approaching the asymptote $\vp^{(d-2)}_\text{inst}(r)$
at $K\to\infty$.

Hence, with increasing the deviation from the gap edge
into the classically forbidden region,
one gradually crosses over from the Larkin-Ovchinnikov
to the Meyer-Simons instanton. The crossover takes place
at the dimensionless energy $\eps_*$ given by Eq.~(\ref{eps*}).
Equivalently the latter may be estimated just by equating
the results (\ref{subgap-LO}) and (\ref{subgap-MS}).

\section{Random order parameter model: Summary and applications}
\label{S:ROP-summary}

Having established the replica structure of the instantons
in the ROP model
we now comment on the limits of validity of the above analysis
and consider various applications of the ROP model
in a more general context.

\subsection{Validity of the instanton analysis}
\label{SS:L}

Our analysis was based on two implicit assumptions:
(i) large instanton action allowing to use the saddle-point approximation, and
(ii) small deviation from the gap edge, $\eps\ll1$, allowing to expand
Eq.~(\ref{F(psi)-def}) and get the universal system (\ref{SP12}).
Once we know the resulting expressions (\ref{subgap-LO}) and (\ref{subgap-MS})
for the DOS tails (at $\eps\ll\eps_*$ and $\eps\gg\eps_*$, respectively),
we can verify these assumptions a posteriori.
Depending on the relations between the parameters of the ROP model
one can identify the following regimes:
\begin{enumerate}
\item[I.]
$\ff(0)/\Delta_0^2\xi^d < (\eta^{2/3}/g_\xi)^{(8-d)/(6-d)}$.

The tail is described by Eq.~(\ref{subgap-MS}) for all $\eps$.
The Larkin-Ovchinnikov tail does not exist
since the corresponding action is smaller than 1.

\item[IIA.]
$(\eta^{2/3}/g_\xi)^{(8-d)/(6-d)} < \ff(0)/\Delta_0^2\xi^d < \eta^{2/3}/g_\xi$.

The main part of the tail (for small $\eps<\eps_*$)
follows the Larkin-Ovchinnikov result (\ref{subgap-LO}).
The tail (\ref{subgap-MS}) exists at large $\eps>\eps_*$
where it is extremely small.

\item[IIB.]
$\eta^{2/3}/g_\xi < \ff(0)/\Delta_0^2\xi^d$.

The tail follows the Larkin-Ovchinnikov result (\ref{subgap-LO}) for all $\eps$.
The tail (\ref{subgap-MS}) does not exist.
\end{enumerate}

In each of the regimes only one type of the instantons is important.
The other one either does not exist or is unobservable.

\subsection{Role of dimensionality}
\label{SS:dim}

Larkin and Ovchinnikov approach \cite{LO71} to the ROP model is based
on the idea of separation of scales: short-scale fluctuations of $\Delta(\br)$
produce an effective depairing $\eta$ leading to the formation
of the AG-like hard gap, whereas long-scale fluctuations
are responsible for the gap smearing. Due to the presence of
the diffusive pole in the integrand in Eq.~(\ref{eta}) for $\eta$,
the possibility of such separation depends on the dimensionality of the problem.

In 3D, the integral in Eq.~(\ref{eta}) is determined by large
momenta, $q\sim r_c^{-1}$, where $r_c$ is the correlation length
of the fluctuating field $\Delta_1(\br)$, leading to the estimate
\be
\label{eta-3D}
  \eta_\text{3D}
  =
  \frac{1}{2\pi^2\Delta_0^2\xi^2}
  \int_0^\infty \ff(q) dq
  \sim
  \frac{\ff(0)}{2\pi^2\Delta_0^2\xi^2r_c}
  .
\ee
The long-wavelength theory (\ref{S3}) is then applicable already
for scales $r\gtrsim r_c$.

The 2D case is marginal since the integral in Eq.~(\ref{eta}) is logarithmic.
Its upper cutoff is again given by $r_c^{-1}$, whereas
the lower limit must be determined with care.
According to Ref.~\cite{FS2012}, with logarithmic accuracy
an appropriate cutoff is established by replacing
$Dq^2\mapsto Dq^2 + D/L_E^2$,
where the length $L_E$ is defined in Eq.~(\ref{LE}).
As a result, the depairing factor becomes energy-dependent:
\be
\label{eta-2D}
  \eta_\text{2D}(E)
  =
  \frac{1}{2\pi}
  \frac{f(0)}{\Delta_0^2\xi^2}
  \ln\frac{\min(L_E,L_g)}{r_c}
  ,
\ee
where we had to introduce an infrared length scale $L_g$
in order to regularize the otherwise divergent $\eta_\text{2D}(E\to E_g)$.
Its appearance is related to the breakdown
of the mean-field approximation in the narrow region
$|E-E_g|\lesssim \Gamma_\text{tail}$, where proliferation
of instantons generates a finite correlation length
$L_g \sim \xi (E_g/\Gamma_\text{tail})^{1/4}$.
Here one should use the Meyer-Simons (Larkin-Ovchinnikov)
expression for $\Gamma_\text{tail}$ provided the regime
I (II) is realized [see discussion in Sec.~\ref{SS:L}].

In 1D, the integral (\ref{eta}) is divergent in the infrared,
indicating that analytical treatment based on the idea of scale
separation is not possible, and $\Gamma_\text{tail} \approx \Gamma$.

\subsection{Applications of the ROP model}
\label{SS:ROP-apps}

\subsubsection{Random coupling constant model}

This is the model initially considered by Larkin and Ovchinnikov \cite{LO71}.
It can be reduced to the ROP model in the following way.
A fluctuating Cooper constant introduces quenched fluctuations
in the order parameter field which should be determined from
the self-consistency equation. The latter can be written
in the Matsubara representation and solved perturbatively.
In the linear order one gets \cite{LO71}
\be
\label{Delta-lambda}
  \lambda_0^{-1} \Delta_1(\bq)
  +
  \delta\lambda^{-1}(\bq)\Delta_0
  =
  \pi T \sum_\epsilon \frac{\partial F(\epsilon,\bq)}{\partial\Delta(\bq)}
  \Delta_1(\bq)
  ,
\ee
where $F(\epsilon,\bq)$ is the Fourier transform of the quasiclassical
Gor'kov function $F(\epsilon,\br)=\sin\theta(\epsilon,\br)$, and its derivative
with respect to $\Delta_1$ can be obtained from the Usadel equation.
Solving for $\Delta_1(\bq)$ we arrive at the linear relation
\be
  \Delta_1(\bq)
  =
  -
  \Delta_0 L_0(q) \, \delta\lambda^{-1}(\bq) ,
\ee
where $L_0(q)$ is the static propagator of superconducting fluctuations
in the BCS theory:
\be
  L_0^{-1}(q)
  =
  \pi T \sum_\epsilon
  \frac{\Delta_0^2+\mathfrak{E}(\epsilon)Dq^2/2}
  {\mathfrak{E}^2(\epsilon) [\mathfrak{E}(\epsilon)+Dq^2/2]}
  ,
\ee
and
\be
\label{E(e)}
  \mathfrak{E}(\epsilon) = \sqrt{\epsilon^2+\Delta_0^2} .
\ee
At zero temperature integration over Matsubara energies $\epsilon$
can be performed analytically \cite{LO71,MeyerSimons2001},
leading to:
\begin{multline}
\label{L(q)}
  L_0(q)
  =
  \frac{2\tilde q^2}
  {
    \pi
    - 4 \sqrt{1-\tilde q^4} \arctan\sqrt{\frac{1-\tilde q^2}{1+\tilde q^2}}
  }
\\{}
  =
  \begin{cases}
    1 - \pi \tilde q^2/4 + \dots, & \tilde q\ll1; \\[6pt]
    1/\ln \tilde q^2 + \dots, & \tilde q\gg1;
  \end{cases}
\end{multline}
where $\tilde q=q\xi$.

Thus we end up with the ROP model specified by the correlation
function
\be
  f(q)
  \equiv
  \corr{\Delta_1 \Delta_1}_{\bq}
  =
  \Delta_0^2 L^2(q)
  \corr{\delta\lambda^{-1} \delta\lambda^{-1}}_{\bq}
  .
\ee

The random coupling constant model can be mapped onto the ROP model
as long as fluctuations of $\lambda(\br)$ are weak
(the resulting depairing parameter $\eta\ll1$). Otherwise
it is not sufficient to use the first-order perturbation theory
in Eq.~(\ref{Delta-lambda}).

\subsubsection{Gap fluctuations in NS systems}

The simplest setup where disorder leads to formation of the subgap states
is the model of the NS junction \cite{Vavilov2001,OSF01}.
Here superconductive correlations are induced into the normal metal
due to the proximity effect, opening the (mini)gap in the excitation
spectrum \cite{GKI04}.
In a long diffusive junction (with size $L\gg\xi$)
a minigap is of the order of the Thouless energy: $E_g\sim\ETh=D/L^2$
\cite{ZhouCharlat,Taras-Semchuk-Altland}.
In the normal part of the junction the order parameter is absent
and the only source of disorder is due to random position of potential
impurities. These \emph{mesoscopic fluctuations}\ are known to be responsible
for various sample-to-sample fluctuations such as universal
conductance fluctuations \cite{UCF}, mesoscopic fluctuations of the
Josephson current \cite{altshuler87,Manuel}, etc.\ \cite{MesoPhenomenaInSolids}.

The DOS tail due to mesoscopic fluctuations in NS hybrid systems
\cite{Vavilov2001,OSF01} follows the result (\ref{subgap-MS}).
This sheds light on the physical origin of the instanton of the action
$S_0+S_\eta$ discussed in Sec.~\ref{SS:MS}: it describes DOS
smearing due to mesoscopic fluctuations of the quasiparticle response
to a constant (for this particular problem, zero) order parameter.
In other words, randomness of impurities' positions translates into
fluctuations of the quasiparticle Green function $Q(\br)$.

\subsubsection{Mesoscopic fluctuations of the order parameter}
\label{SSS:MF}

In disordered superconductors, mesoscopic fluctuations lead to fluctuations
of the order parameter \cite{FS2012,SF2005}.
The latter are generated by fluctuations of $Q(\br)$
if one takes the self-consistency equation into account.

In the 2D case and in the presence of the Coulomb interaction,
the order parameter correlation function was calculated in Ref.~\cite{FS2012}.
The correlation length of these fluctuations is of the order
of the zero-temperature coherence length, $r_c\sim\xi(0)$.
Their magnitude at $T=0$ and $q=0$ are given by
\be
\label{f(0)-FS-0}
  \frac{f_\text{2D}^\text{MF+Coulomb}(0)}{\Delta_0^2\xi^2}
  =
  \frac{2\pi}{g(g-g_c)}
  ,
\ee
where $g_c = \ln^2(\hbar/T_{c0}\tau)/2\pi$ is the critical conductance
for the fermionic mechanism of the superconductivity suppression~\cite{Finkelstein}
($T_{c0}$ is the transition temperature in the clean system and
$\tau$ is the elastic scattering time).

Evaluating the depairing parameter from Eq.~(\ref{eta-2D}),
we find that the regime IIB is always realized [see Sec.~\ref{SS:L}].
Therefore gap smearing in homogeneously disordered superconducting
films is always described by the Larkin-Ovchinnikov
mechanism leading to Eq.~(\ref{subgap-LO}),
and there is no room for the Meyer-Simons instanton \cite{FS2012}.

For completeness we present here the general expression for the correlation
function of the order parameter due to mesoscopic fluctuations at arbitrary
space dimensionality $d$ but in the absence of the Coulomb effects:
\be
\label{f(q)-MF}
  {f^\text{MF}(q)}
  =
  \frac{L_0^2(q)\Delta_0^2T^2 }{\nu^2}
  \sum_{\epsilon,\epsilon'}
  \int
  \frac{\Pi_{\epsilon\epsilon'}(k) \Pi_{\epsilon\epsilon'}(k-q)}
  {\mathfrak{E}(\epsilon)\mathfrak{E}(\epsilon')}
  \frac{d^d\bk}{(2\pi)^d}
  ,
\ee
where $\mathfrak{E}(\epsilon)$ is defined in Eq.~(\ref{E(e)}),
and
$
  \Pi_{\epsilon\epsilon'}(k)
  =
  [Dk^2 + \mathfrak{E}(\epsilon) + \mathfrak{E}(\epsilon')]^{-1}
$
is the diffusion propagator on top of the BCS state.
In particular, at $T=0$ and $q=0$ one gets
\be
\label{f(0)-MF-0}
  \frac{f^\text{MF}(0)}{\Delta_0^2\xi^d}
  \sim
  \frac{1}{g_\xi^2}
  .
\ee
Since mesoscopic fluctuations are inevitably present in any disordered system,
Eq.~(\ref{f(0)-MF-0}) is the lower bound for the order parameter fluctuations
in disordered superconductors.

\section{Magnetic impurities}
\label{S:magnetic}

\subsection{Abrikosov-Gor'kov model}

Now we turn to the situation when the BCS coherence peak is smeared
by magnetic disorder. We restrict ourselves to the AG model \cite{AG}
of Gaussian point-like magnetic impurities (a more general case will
be considered elsewhere \cite{Yasha-2B}) specified by the correlation
function of the exchange field:
\be
\label{<hh>}
  \corr{h_i(\br) h_j(\br')}
  =
  \frac{\delta_{ij} \, \delta(\br-\br')}{6\pi\nu\tau_s} ,
\ee
where $\tau_s$ is the spin-flip scattering time.
The latter plays the role of the pair-breaking time which determines
the pair-breaking parameter
\be
\label{eta-AG}
  \eta = \frac{1}{\tau_s\Delta_0}
  .
\ee
The vector $\mathbf h$ is three-dimensional,
while the effective dimensionality of the sample may be different.

Formation of the subgap states in a superconductor with weak magnetic impurities
was studied by Lamacraft and Simons \cite{LamacraftSimons}
who came to the result essentially coinciding with Eq.~(\ref{subgap-MS}).
However, inspired by the preceding analysis of the ROP model one may
expect that there should be instantons related to optimal fluctuations of
the exchange field $\mathbf h$ and/or order parameter field $\Delta(\br)$,
leading to the Larkin-Ovchinnikov tail (\ref{subgap-LO}).
Below we study this instanton contribution and demonstrate that the main
part of the subgap DOS tail may be described either by the
Lamacraft-Simons or by the Larkin-Ovchinnikov results,
depending of the values of $g_\xi$ and $\eta$ [see Sec.~\ref{SS:AG-results}].
In the case when the tail is due to the Larkin-Ovchinnikov optimal fluctuation,
it arises as a result of mesoscopic fluctuations of the order parameter.

\subsection{Sigma-model action}

We use the real-energy replica sigma model introduced in Sec.~\ref{SS:RSM}.
Before averaging over magnetic disorder the initial action expanded
to the second order in the impurity magnetization $\mathbf{h}(\br)$
takes the form \cite{we-spinflip}:
\be
\label{S012}
  S = S_0 + S_1 + S_2 ,
\ee
where $S_0$ is the action for the uniform superconductor given
by Eq.~(\ref{S0}), and the terms $S_{1,2}$ describe magnetic impurities:
\begin{gather}
\label{Sh}
  S_1
  =
  -i \pi\nu
  \int d\br \,
  \mathbf{h}(\br)
  \tr (\tau_3 \bm{\sigma} Q)
  ,
\\
\label{Shh}
  S_2
  =
  - \frac{(\pi\nu)^2}2 \int d\br \, h_i(\br) h_j(\br')
  \tr \bigl( \tau_3 \sigma_i Q \tau_3 \sigma_j Q \bigr)
  ,
\end{gather}
where $\sigma_i$ are Pauli matrices in the spin space.

Averaging over $\mathbf{h}$ with the correlation function (\ref{<hh>})
generates two terms, $S_\eta=\corr{S_2}$ and $S_\text{dis}=-\corr{S_1^2}/2$,
with different structures in the replica space:
\begin{gather}
\label{Seta-AG}
  S_\eta
  =
  - \frac{\pi\nu\Delta_0\eta}{12}
  \int d\br \,
  \tr ( \tau_3 \bm{\sigma} Q )^2
  ,
\\
\label{Sdis-AG}
  S_\text{dis}
  =
  \frac{\pi\nu\Delta_0\eta}{12}
  \int d\br \,
  (\tr \tau_3 \bm{\sigma} Q)^2 ,
\end{gather}
where the depairing parameter $\eta$ is given by Eq.~(\ref{eta-AG}).

As a result, the effective action describing gap fluctuations
in the presence of a Gaussian short-range magnetic disorder takes the form:
\be
\label{S3-AG}
  S = S_0 + S_\eta + S_\text{dis} .
\ee
The structure of the terms $S_\eta$ [Eq.~(\ref{Seta-AG})]
and $S_\text{dis}$ [Eq.~(\ref{Sdis-AG})] is pretty similar
to that of the analogous terms, (\ref{Seta}) and (\ref{Sdis}),
in the ROP model. Note that contrary to the ROP model,
the depairing term $S_\eta$ is generated automatically
after averaging over $\delta$-correlated magnetic disorder.

In the analysis of the action (\ref{S3-AG}), Lamacraft and Simons
\cite{LamacraftSimons} considered only singlet configurations
of the field $Q(\br)$. Then the term $S_\text{dis}$ can be discarded,
while the term $S_\eta$ just coincides with
the analogous term (\ref{Seta}) in the ROP model.
In the absence of a field responsible for optimal fluctuations,
the authors of Ref.~\cite{LamacraftSimons}
reproduced the result (\ref{subgap-MS}).

\subsection{Effective fluctuators in the singlet sector}

In order to go beyond the analysis of Ref.~\onlinecite{LamacraftSimons}
one has to identify an effective fluctuator in the singlet sector
which might be responsible for the Larkin-Ovchinnikov optimal fluctuation
at $\eps\ll\eps_*$ ($K\to\infty$).
We focus on the singlet sector since it becomes massless at $E\to E_g$,
whereas the triplet is not [see Eqs.~(\ref{Us2-lin}) and (\ref{rs})].
Therefore the instanton solution with the stretched-exponent action
of the type (\ref{rho-Gtail}) may arise only in the singlet component of $Q(\br)$.

There are several sources of fluctuations in the singlet sector:
\begin{enumerate}
\item[(i)]
Mesoscopic fluctuations of the order parameter with the correlation function
given by Eq.~(\ref{f(0)-MF-0}). They arise due to fluctuations of potential
impurities and are insensitive to weak magnetic disorder.

\item[(ii)]
Fluctuations in the singlet component of the Green function $Q(\br)$
generated through its triplet component due to nonlinearity
of the Usadel equation (referred to as direct fluctuations).
These fluctuations can be described in terms of an effective
order parameter field $\Delta_1^{(\Phi)}(\br)$.
Its correlation function is calculated in Appendix.
In the limit $T=0$ and $q=0$ it can be estimated as
\be
\label{f-ii}
  \frac{\ff^{(\Phi)}(0)}{\Delta_0^2\xi^d}
  \sim
  \frac{\eta^{(4+d)/6}}{g_\xi^2}
  .
\ee

\item[(iii)]
Fluctuations of the order parameter due to randomness in $\mathbf h$
(referred to as indirect fluctuations) calculated in Appendix.
In the limit $T=0$ and $q=0$ the corresponding correlation function
can be estimated as
\be
\label{f-iii}
  \frac{\ff(0)}{\Delta_0^2\xi^d}
  \sim
  \frac{\eta^2}{g_\xi^2} .
\ee

\end{enumerate}
The presence of the the factors $g_\xi^2$ in the denominators
of Eqs.~(\ref{f-ii}) and (\ref{f-iii}) can be easily explained.
Due to the vector structure of the random field $\mathbf h(\br)$,
an effective fluctuator in the singlet sector will be
proportional to $\mathbf{h}^2$, with its variance, $f$, scaling
as $\corr{\mathbf{h}^2}^2\propto\nu^{-2}\propto g_\xi^{-2}$.

\subsection{Result}
\label{SS:AG-results}

Comparing Eqs.~(\ref{f(0)-MF-0}), (\ref{f-ii}) and (\ref{f-iii})
we conclude that mesoscopic fluctuations of the order parameter
is the leading source of disorder in $\Delta(\mathbf{r})$ for the magnetic impurities model.
Therefore there is a competition of the Larkin-Ovchinnikov
result (\ref{subgap-LO}) with $f(0)$ given by Eq.~(\ref{f(0)-MF-0})
and the Lamacraft-Simons dependence (\ref{subgap-MS}).
According to Sec.~\ref{SS:L}, the winner depends on the values
of $g_\xi$ and $\eta$:
\begin{itemize}
\item
Regime I is realized for $g_\xi > \eta^{-\frac{2(8-d)}{3(4-d)}}$.
The subgap DOS follows Eq.~(\ref{subgap-MS}).
\item
Regime II is realized for $g_\xi < \eta^{-\frac{2(8-d)}{3(4-d)}}$.
The subgap DOS follows Eq.~(\ref{subgap-LO}).
\end{itemize}

\section{Conclusion}
\label{S:discussion}

This work was motivated by the discrepancy of the two instanton
approaches to the problem of the subgap states in disordered superconductors.
We have analyzed the replica structure of a generic instanton solution
and demonstrated that the instanton of Larkin and Ovchinnikov \cite{LO71}
can be continuously deformed to the instanton of Simons and others
\cite{MeyerSimons2001,LamacraftSimons,MarchettiSimons}
with decreasing the energy into the classically forbidden region.

Existence of two different instanton types
is related to the presence of two types of disorder in the system:
(i) the potential disorder responsible for diffusive
motion of electrons and (ii) extra randomness in the some other
characteristics of the sample, e.g., the Cooper coupling constant,
the order parameter field, random spin exchange field, etc.
In the quasiclassical theory of dirty superconductors,
the potential (type-i) disorder is averaged out in the very beginning.
The resulting Usadel equations are nonlinear already in the
absence of type-ii disorder. Averaging over the latter
brings an additional nonlinearity, which competes with the intrinsic
nonlinearity of the problem.
The relative strength of the two nonlinear terms is controlled by the
proximity to the gap edge.
For $\eps\ll\eps_*$, intrinsic nonlinearity is not important
and the situation is similar to the problem of the linear Schr\"odinger
equation with disorder. The instanton then corresponds to the optimal
fluctuation of the random field, yielding the Larkin-Ovchinnikov
result (\ref{subgap-LO}).
In the opposite limit, $\eps\gg\eps_*$, only intrinsic nonlinearity
of the problem is relevant. The nonlinear equations of motion
still allow a bounce solution corresponding to the instanton
of Simons and others.

Physically, the instanton of Larkin and Ovchinnikov describes
an optimal fluctuation of the order parameter field which decreases
the local value of the gap. The instanton of Simons and others
describes mesoscopic fluctuations of quasiparticle response
at a fixed value of the order parameter.

Depending on the parameters of the problem it might happen that
the instanton action at the crossover energy, $\eps\sim\eps_*$,
is smaller that 1. In this situation, the Larkin-Ovchinnikov
instanton does not exist and the density of the subgap states
is described by Eq.~(\ref{subgap-MS}).
Otherwise the main part of the tail is described by Eq.~(\ref{subgap-LO}),
while its far asymptotics (\ref{subgap-MS}) is practically unobservable.

This general structure of the subgap DOS tail is analyzed for a
number of superconducting problems with disorder.
In particular, we reconsidered the gap smearing in the Abrikosov-Gor'kov
model of weak paramagnetic impurities \cite{LamacraftSimons} and showed that depending
on the parameters of the problem the DOS tail is described
either by Eq.~(\ref{subgap-LO}) or by Eq.~(\ref{subgap-MS}).

Finally, we emphasize that our analysis applies to dirty superconductors
and NS hybrids with diffusive electron dynamics described by the Usadel
equation. Much less is known on the nature of the proximity gap in ballistic
chaotic systems which is determined by the competition of the mean free time,
the Ehrenfest time, and the escape time \cite{belzig,AT,VL2003}.
Going beyond the mean-field analysis and generalizing our findings
to that type of systems remains an open problem.

\acknowledgments

We thank Ya.\ V. Fominov and S. E. Korshunov for useful discussions.
This work was partially supported by the Russian Ministry of Education
and Science (Contract No.\ 8678), the program ``Quantum mesoscopic
and disordered structures'' of the RAS, and RFBR grant No.\ 13-02-01389.

\appendix

\section{Disorder in the singlet sector due to magnetic impurities}

In this Appendix we calculate the correlation functions of effective
disorder in the singlet sector due to randomness in $\mathbf h(\br)$.

\subsection{Triplet Usadel equation}

Induced magnetization can be described with the help of the triplet
Usadel equations. Following Ref.~\onlinecite{IF2006}
we parametrize the $Q$ matrix in terms of the spectral angle
$\theta$ and the magnetization vector $\mathbf M$ as
\begin{multline}
\label{Q-param}
  Q_0
  =
  M_0 \sigma_0
  (\tau_3 \cos \theta + \tau_1 \sin \theta)
\\{}
+ i \mathbf M \bm{\sigma}
  (\tau_3 \sin \theta - \tau_1 \cos \theta) ,
\end{multline}
where $M_0=\sqrt{1+\mathbf M^2}$.
The resulting equations for the singlet ($\theta$) and triplet
(${\mathbf M}$) components in the Matsubara representation
have the form \cite{we-spinflip}:
\begin{subequations}
\begin{multline}
  \frac D2 \nabla^2 \theta
+ M_0 \left( -\epsilon \sin\theta + \Delta_0 \cos\theta \right)
- ( \mathbf{h M} ) \cos\theta
\\
- \Delta_0\eta \left( 1 + \frac{2}{3} \mathbf M^2 \right) \sin \theta \cos\theta
= 0 ,
\label{Us1}
\end{multline}
\vskip-15pt
\begin{multline}
\frac D2 \left( \mathbf M \nabla^2 M_0 - M_0 \nabla^2 \mathbf M \right)
+ \mathbf M ( \epsilon \cos\theta + \Delta_0 \sin\theta )
\\
- M_0 \mathbf h \sin\theta_0 + \frac{1}{3} \Delta_0\eta M_0 \mathbf M  \cos 2\theta
= 0 .
\label{Us2}
\end{multline}
\end{subequations}
In the absence of $\mathbf h$, we have $\mathbf M=0$
and the spectral angle $\theta_0(\epsilon)$ should be obtained from
the AG equation (\ref{Usadel-AG}) analytically continued to Matsubara
energies, $iE\to-\epsilon$.

The linear response of the magnetization $\mathbf M$ to the field $\mathbf h$
can be found from the triplet equation (\ref{Us2}), which yields in
the momentum representation:
\be
  \mathbf M(\bq)
  =
  \frac{\sin\theta_0}{\xi^2 q^2 + \mu_t(\epsilon)}
  \frac{\mathbf h(\bq)}{\Delta_0}
  ,
\label{Us2-lin}
\ee
where
\be
\label{mu-t}
  \mu_t(\epsilon)
  =
  \frac{\epsilon}{\Delta_0} \cos\theta_0(\epsilon)
  + \sin\theta_0(\epsilon) + \frac{\eta}{3} \cos2\theta_0(\epsilon)
\ee
has the meaning of a mass of the triplet modes.

Now expanding the singlet equation (\ref{Us1})
to the second order in $\mathbf h$ and using
Eq.~(\ref{Usadel-AG}) we obtain
\be
\label{singlet-Phi}
- \xi^2 \nabla^2 \theta
+ (\epsilon/\Delta_0) \sin\theta - \cos\theta
+ \eta \sin \theta \cos\theta
=
  \Phi(\br) ,
\ee
where $\Phi_\epsilon(\br)$ acts as an effective source of singlet fluctuations:
\be
  \Phi_\epsilon(\br)
  =
- \frac{\eta\mathbf M^2}{6} \sin \theta_0(\epsilon) \cos\theta_0(\epsilon)
- \frac{\mathbf{h M}}{\Delta_0} \cos\theta_0(\epsilon)
  .
\label{Phi}
\ee

A nonzero average $\corr{\Phi_\epsilon(\br)}$
leads to renormalization of $\Delta_0$ and $\eta$,
while the strength of disorder in the singlet sector
is determined by the irreducible correlator
\be
\label{<PhiPhi>}
  \ccorr{\Phi_\epsilon\Phi_{\epsilon'}}_{\bq}
  =
  \frac{\eta^2 \sin 2\theta_0 \sin 2\theta_0'}{96(\pi\nu\Delta_0)^2}
  \int
  Z_{\epsilon}(k,q)
  Z_{\epsilon'}(k,q)
  \frac{d^d\bk}{(2\pi)^d}
  ,
\ee
where $\theta_0=\theta_0(\epsilon)$, $\theta_0'=\theta_0(\epsilon')$,
\be
  Z_{\epsilon}(k,q)
  =
    \Pi_\epsilon(k) + \Pi_\epsilon(k-q)
  +
    \frac{\eta\sin^2\theta_0(\epsilon)}{3} \Pi_\epsilon(k) \Pi_\epsilon(k-q)
  ,
\ee
and
\be
  \Pi_\epsilon(q) = \frac{1}{\xi^2 q^2 + \mu_t(\epsilon)}
\ee
is the triplet diffusion propagator on top of the AG state.

Further analysis goes differently for direct and indirect
fluctuations of $\Phi(\br)$.

\subsection{Direct fluctuations of $\Phi$}

According to Eq.~(\ref{singlet-Phi}), behavior of quasiparticles with energy $\epsilon$
in the field of a fluctuating $\Phi_{\epsilon}(\br)$ and constant $\Delta(\br)=\Delta_0$
is formally equivalent to that in the field of a fluctuating order parameter with
\be
\label{DeltaPhi}
  \Delta_1^{(\Phi)}(\br)
  =
  \frac{\Delta_0 \Phi_{\epsilon}(\br)}{\cos\theta_0(\epsilon)}
  .
\ee
For the problem of the DOS tail we need real energies near the gap edge,
$\epsilon=-iE\to -iE_g$. Since the triplet sector remains massive
at the edge, we may simply evaluate $\Phi_\eps$ right at $\epsilon=-iE_g$,
when
\be
\label{rs}
  \mu_t(-iE_g)
  \equiv
  \xi^2r_s^{-2}
  =
  \frac{4}{3} \eta^{1/3} \left( 1-\frac12\eta^{2/3} \right)
  .
\ee
Here $r_s$ is the spin-rigidity length at the gap edge,
which is finite in contrast to a divergent length $L_E$
in the singlet sector [Eq.~(\ref{LE})].

Fluctuations of the field $\Delta_1^{(\Phi)}(\br)$
are characterized by the irreducible correlator
$\ff^{(\Phi)}(\bq) = \ccorr{\Delta_1^{(\Phi)}\Delta_1^{(\Phi)}}_\bq$
which can be extracted from Eqs.~(\ref{cosh-g}), (\ref{<PhiPhi>}) and (\ref{DeltaPhi}):
\be
  \ff^{(\Phi)}(q)
  =
  \frac{\eta^{4/3}}{24(\pi\nu)^2}
  \int
  Z_{-iE_g}^2(k,q)
  \frac{d^d\bk}{(2\pi)^d} ,
\ee
where in calculating $Z_{-iE_g}(k,q)$ it should be taken into account
that $\sin\theta_0(-iE_g)=\cosh\psi_g=\eta^{-1/3}$.
The correlation length of the field $\Delta_1^{(\Phi)}(\br)$
is of the order of $r_s$, and the zero-momentum correlation
function can be estimated as
\be
\label{f(0)-direct}
  \frac{\ff^{(\Phi)}(0)}{\Delta_0^2\xi^d}
  \sim
  \frac{\eta^{4/3}}{g_\xi^2}
  \left( \frac{r_s}{\xi} \right)^{4-d}
  ,
\ee
leading to Eq.~(\ref{f-ii}).

\subsection{Indirect fluctuations of $\Phi$}

The field $\Phi_\epsilon(\br)$ also affects quasiparticle behavior
indirectly by inducing quenched inhomogeneity in the order parameter
field. Fluctuations of $\Delta(\br)$ can be obtained from
the linearized self-consistency equation
[compare with Eq.~(\ref{Delta-lambda})]:
\be
\label{Delta-h}
  \lambda^{-1} \Delta_1(\bq)
  =
  \pi T \sum_\epsilon
  \left[
  \frac{\partial F(\epsilon,\bq)}{\partial\Delta(\bq)}
  \Delta_1(\bq)
  + \delta F(\epsilon,\bq)
  \right]
  ,
\ee
where $\delta F(\epsilon,\bq) = \delta\sin\theta(\epsilon,\bq)$
is a fluctuating part of the anomalous Matsubara Green function
evaluated at a constant $\Delta(\br)=\Delta_0$.
Solving for $\Delta_1(\bq)$ we get
\be
\label{Delta-h2}
  \Delta_1(\bq)
  =
  L_0(q) \,
  \pi T \sum_\epsilon
  \delta F(\epsilon,\bq)
  ,
\ee
where $L_0(q)$ is the static fluctuation propagator on top of the AG
solution with a finite $\eta$ [note that Eq.~(\ref{L(q)}) refers
to the BCS case with $\eta=0$]. It is given by
\be
\label{L(q)2}
  L_0^{-1}(q)
  =
  \frac{\pi T}{\Delta_0} \sum_\epsilon
  \left[
    \sin\theta_0(\epsilon)
    -
    \frac{\cos^2\theta_0(\epsilon)}
    {\xi^2 q^2 + \mu_s(\epsilon)}
  \right]
  ,
\ee
where $\mu_s(\epsilon)$ is the mass of the singlet modes
[compare with Eq.~(\ref{mu-t})]:
\be
\label{mu-s}
  \mu_s(\epsilon)
  =
  \frac{\epsilon}{\Delta_0} \cos\theta_0(\epsilon)
  + \sin\theta_0(\epsilon) + \eta \cos2\theta_0(\epsilon) .
\ee
In the non-magnetic case ($\eta=0$), Eq.~(\ref{L(q)2}) reduces to Eq.~(\ref{L(q)}).

The correction $\delta F(\epsilon,\bq) = \cos\theta_0(\epsilon) \, \delta\theta(\epsilon,\bq)$
induced by magnetic disorder follows from Eq.~(\ref{singlet-Phi}):
\be
  \delta F(\epsilon,\bq)
=
  \frac{\cos\theta_0(\epsilon)\,\Phi_\epsilon(\bq)}{\xi^2 q^2 + \mu_s(\epsilon)} .
\ee
For the irreducible correlator
$\ff(\bq) = \ccorr{\Delta_1\Delta_1}_\bq$ we get:
\be
  \ff(q)
  =
  L_0^2(q) (\pi T)^2 \sum_{\epsilon,\epsilon'}
  \frac{\cos\theta_0(\epsilon)\cos\theta_0(\epsilon')\ccorr{\Phi_\epsilon\Phi_{\epsilon'}}_\bq}
  {[\xi^2 q^2 + \mu_s(\epsilon)][\xi^2 q^2 + \mu_s(\epsilon')]} .
\ee

The field $\Delta_1(\br)$ is correlated at the scale of the zero-temperature
coherence length, $r_c\sim\xi(0)$, and its correlation function
at $T=0$ and $q=0$ is given by Eq.~(\ref{f-iii}).

Note that in the limit $\eta\ll1$ the correlation function of indirect fluctuations
is much smaller than the correlation function of direct fluctuations
[Eq.~(\ref{f(0)-direct})]. This is a consequence of the fact
that $\Delta_1^{(\Phi)}$ accumulates fluctuations from the region of the size
of the spin length $r_s$, whereas $\Delta_1$ accumulates fluctuations
from the much smaller region of the size of the coherence length $\xi$.


\begin{thebibliography}{99}

\bibitem{AG1958}
A. A. Abrikosov and L. P. Gor'kov,
Zh. Eksp. Teor. Fiz. {\bf 35}, 1558 (1958); {\bf 36}, 319 (1959)
[Sov. Phys. JETP {\bf 8}, 1090 (1959); {\bf 9}, 220 (1959)].

\bibitem{Anderson1959}
P. W. Anderson, J. Phys. Chem. Solids {\bf 11}, 26 (1959).

\bibitem{AG}
A. A. Abrikosov and L. P. Gor'kov,
Zh. Eksp. Teor. Fiz. \textbf{39}, 1781 (1960)
[Sov. Phys. JETP \textbf{12}, 1243 (1961)].

\bibitem{Anthore-2003}
A. Anthore, H. Pothier, and D. Esteve,
Phys. Rev. Lett. \textbf{90}, 127001 (2003).

\bibitem{Maki-H}
K. Maki,
Prog. Teor. Phys. (Kyoto) \textbf{29}, 333 (1963);
\textbf{31}, 731 (1964).

\bibitem{Maki-Superconductivity}
K. Maki in \emph{Superconductivity}, edited by
R. D. Parks (Marcel Dekker, New York, 1969), p. 1035.

\bibitem{LO71}
A. I. Larkin and Yu. N. Ovchinnikov,
Zh. Eksp. Teor. Fiz. \textbf{61}, 2147 (1971)
[Sov. Phys. JETP \textbf{34}, 1144 (1972)].

\bibitem{ZL}
J. Zittartz and J. S. Langer,
Phys. Rev. \textbf{148}, 741 (1966).

\bibitem{Lif}
I. M. Lifshitz,
Zh. Eksp. Teor. Fiz. \textbf{53}, 743 (1968)
[Sov. Phys. JETP \textbf{26}, 462 (1972)].

\bibitem{Usadel}
K. Usadel, Phys. Rev. Lett. \textbf{25}, 507 (1970).

\bibitem{com-d}
In Ref.~\onlinecite{LO71} only the 3D case with weak disorder, $\eta\ll1$,
was considered, but generalization of this result to arbitrary
dimensionality $d$ is straightforward.

\bibitem{MeyerSimons2001}
J. S. Meyer and B. D. Simons,
Phys. Rev. B \textbf{64}, 134516 (2001).

\bibitem{LamacraftSimons}
A. Lamacraft and B. D. Simons,
Phys. Rev. Lett. \textbf{85}, 4783 (2000);
Phys. Rev. B \textbf{64}, 014514 (2001).

\bibitem{Vavilov2001}
M. G. Vavilov, P. W. Brouwer, V. Ambegaokar, and C. W. J. Beenakker,
Phys. Rev. Lett. \textbf{86}, 874 (2001).

\bibitem{OSF01}
P. M. Ostrovsky, M. A. Skvortsov, M. V. Feigel'man,
Phys. Rev. Lett. \textbf{87}, 027002 (2001).

\bibitem{MarchettiSimons}
F. M. Marchetti and B. D. Simons,
J. Phys. A: Math. Gen. \textbf{35}, 4201 (2002).

\bibitem{TracyWidom}
C. A. Tracy and H. Widom,
Commun. Math. Phys. \textbf{159}, 151 (1994);
\textbf{177}, 727 (1996).

\bibitem{com-rc}
The inequality (\ref{rc<xi}) can be replaced by a weaker
inequality $r_c<L_E$, where $L_E$ is given by Eq.~(\ref{LE}).

\bibitem{FS2012}
M. V. Feigel'man and M. A. Skvortsov,
Phys. Rev. Lett. \textbf{109}, 1470022 (2012).

\bibitem{Efetov}
K. B. Efetov,
\emph{Supersymmetry in Disorder and Chaos}
(Cambridge Univ. Press, Cambridge, 1996).

\bibitem{Finkelstein-review}
A.~M.\ Finkel'stein, in \textit{Soviet Scientific Reviews},
edited by I.~M.\ Khalatnikov (Harwood Academic, London, 1990), Vol.~14.

\bibitem{AST}
A.\ Altland, B.~D.\ Simons, and D.\ Taras-Semchuk,
Adv.\ Phys.\ \textbf{49}, 321 (2000).

\bibitem{we-spinflip}
D. A. Ivanov, Ya. V. Fominov, M. A. Skvortsov, and P. M. Ostrovsky,
Phys. Rev. B \textbf{80}, 134501 (2009).

\bibitem{Silva}
A. Silva and L. B. Ioffe,
Phys. Rev. B \textbf{71}, 104502 (2005).

\bibitem{Parisi-Sourlas-1979}
G. Parisi and N. Sourlas,
Phys. Rev. Lett. \textbf{43}, 744 (1979).

\bibitem{GKI04}
A. A. Golubov, M. Yu. Kupriyanov, and E. Il'ichev,
Rev. Mod. Phys. \textbf{76}, 411 (2004).

\bibitem{Taras-Semchuk-Altland}
D. Taras-Semchuk and A. Altland,
Phys. Rev. B \textbf{64}, 014512 (2001).

\bibitem{ZhouCharlat}
F. Zhou, P. Charlat, B. Spivak, and B. Pannetier,
J. Low Temp. Phys. \textbf{110}, 841 (1998).

\bibitem{UCF}
B. L. Altshuler,
Pis'ma Zh. Eksp. Teor. Fiz. \textbf{41}, 530 (1985)
[Sov. Phys. JETP Lett. \textbf{41}, 648 (1985)];
P. A. Lee and A. D. Stone,
Phys. Rev. Lett. \textbf{55}, 1622 (1985).

\bibitem{altshuler87}
B. L. Altshuler and B. Z. Spivak,
Zh. Eksp. Theor. Fiz. \textbf{92}, 607 (1987)
[Sov. Phys JETP \textbf{65}, 343 (1987)].

\bibitem{Manuel}
M. Houzet and M. A. Skvortsov,
Phys. Rev. B \textbf{77}, 057002 (2008).

\bibitem{MesoPhenomenaInSolids}
{\it Mesoscopic Phenomena in Solids},
Modern Problems in Condensed Matter Sciences Vol. 30,
edited by B. L. Altshuler, P. A. Lee, and R. A. Webb,
(North-Holland, Austerdam, 1991).

\bibitem{SF2005}
M. A. Skvortsov and M. V. Feigel'man,
Phys. Rev. Lett. \textbf{95}, 057002 (2005).

\bibitem{Finkelstein}
A. M. Finkestein, Pis'ma Zh. Eksp. Teor. Fiz. {\bf 45}, 37 (1987)
[Sov. Phys. JETP Lett. {\bf 45}, 46 (1987)];
Physica B {\bf 197}, 636 (1994).

\bibitem{Yasha-2B}
Ya. V. Fominov and M. A. Skvortsov,
in preparation.

\bibitem{belzig}
S. Pilgram, W. Belzig and C. Bruder,
Phys. Rev. B {\bf 62}, 12462 (2000).

\bibitem{AT}
D. Taras-Semchuk and A. Altland,
Phys. Rev. B {\bf 64}, 014512 (2001).

\bibitem{VL2003}
M. G. Vavilov and A. I. Larkin,
Phys. Rev. B {\bf 67}, 115335 (2003).

\bibitem{IF2006}
D. A. Ivanov and Ya. V. Fominov,
Phys. Rev. B \textbf{73}, 214524 (2006).

\end{thebibliography}
\end{document}